# Redefining Influenza Transmission Seasonality Using the Novel Seasonality Index

Branislava Lalic[1], Vladimir Koci[2], Ana Firanj Sremac[1], Zorana Jovanovic Andersen[3]

1 Group for Meteorology and Biophysics, Faculty of Agriculture, University of Novi Sad, 21000 Novi Sad, Serbia
2 Sekspirova 20, 21000 Novi Sad, Serbia
3 Department of Public Health, University of Copenhagen, Oster Farimagsgade 5, 1353 Copenhagen, Denmark

Corresponding author: branislava.lalic@polj.edu.rs

**Abstract**
The impact of climate conditions on influenza epidemiology has mostly been studied by addressing a singular aspect of transmission and a climate variable correlating to it. As climate change unfolds at an unprecedented rate, we urgently need new multidisciplinary approaches that can embrace complexity of disease transmission in the fast-changing environment and help us better understand the implications for health. In this study, we have implemented a novel seasonality index to capture a vast network of climate, infectious, and socio-behavioural mechanisms influencing a seasonal influenza epidemic. We hypothesize that intricate, region-specific behavioural patterns are cross regulating the influenza spreading and dynamics of epidemics with changes in meteorological conditions within a specific season.
To better understand the phenomena, we analysed weekly surveillance data from temperate European countries and redefined seasonal transitions using the seasonality index. This approach allowed us to characterize influenza seasonality more accurately in relation to specific atmospheric conditions. Key findings include: i) a strong correlation between influenza infection rates and the seasonality index across different climate zones and social groups, and ii) a high linear correlation between winter duration, determined by the seasonality index, and the time scale of low-frequency peaks in the infection rates power spectral density.

## 1 Introduction

Throughout history, influenza viruses have caused several devastating pandemics, including the Spanish flu of 1918, which resulted in 20-50 million deaths, the Russian flu, and the 2009 swine flu pandemic, underscoring the global impact of influenza outbreaks. To this day, influenza remains a significant public health concern with over 300,000 annual deaths worldwide attributed to the virus (Tyrrell et al. 1995). Despite achievements in antiviral medications and seasonal vaccinations, the infectivity rate and mortality toll remain high.

The influenza A virus (IAV) is responsible for most severe cases and fatalities, as noted by the center for Disease Control and Prevention (CDC) (Wong et al. 2013). The root of its resilience lays in the virus' ability to evade the immune system through mutations in its surface antigens, particularly hemagglutinin (HA) and neuraminidase (NA), which vary each influenza season (Russell et al. 2021).

Previous studies have demonstrated a correlation between low temperatures in temperate regions and increases in influenza-positive cases (Roussel et al. 2016). In addition to temperature, humidity- both relative and absolute - has been shown to influence viral survival and aerosol transmission, thereby impacting the infection rates (Marr et al. 2019). Nevertheless, there remains a lack of consensus on how key climate factors, which define the climate season itself, are connected to influenza outbreaks, particularly towards the end of the calendar year.

In temperate regions, influenza infection rates typically peak during winter, particularly around the holiday season. During this period, adults experience increased exposure to infected children, who predominantly contract the virus in schools (Lane et al. 2021; Ewing et al. 2016). Additionally, winter conditions - such as low outdoor temperatures, high indoor relative humidity, and increased time spent in enclosed spaces - facilitate viral transmission (Marr et al. 2019). Numerous studies have demonstrated a correlation between meteorological conditions and the spread of influenza in temperate regions (e.g., Shaman et al., 2010; Tamerius et al., 2011). However, identifying a consistent correlation with any single meteorological variable across diverse locations remains challenging. Traditional studies have failed to encompass the full climatological context underlying the concept of a season, especially in the face of climate change. With the ongoing shifts in global atmospheric processes' caused by climate change, it is necessary to challenge conventional concepts of influenza seasonality and its relationship with atmospheric seasons.

We hypothesize that the influenza seasonality is driven by a complex interplay of factors, including the relationship between atmospheric seasonality and fluctuations in infection rates, behavioral patterns influenced by climate seasonality, and the direct effects of climate on the influenza virus. This holistic approach provides a more robust framework for understanding influenza transmission dynamics than focusing on individual meteorological elements. To test this hypothesis, we employ a novel seasonality index, introduced by Lalic et al. (2022) and Lalic and Firanj Sremac (2025), and Power Spectral Density (PSD) analysis to trace influenza seasonality and identify dominant time scales affecting influenza transmission.

*Rationale for using the seasonality index to trace influenza transmission.* To better understand the key drivers of influenza spread across different time scales, we analyzed PSD of infection rate (IR=influenza_positive/tested_individuals*100) time series. PSD is a powerful tool for examining time series data coming from biomedical research. Such time series often represent complex biomedical phenomena governed by processes occurring across various temporal and spatial scales. Even though PSD spectra cannot decipher the exact underlying processes, they effectively reveal the time scales at which these processes occur – focusing on "when" rather than "what". The PSD is a measure that describes how the power of the time series is distributed across different frequencies, helping to identify which frequencies, i.e. time scales, contribute most to the signal's variance. In the case of influenza time series, low-frequency components represent long-term, slow variations associated with seasonality of meteorological conditions, population behaviour, long-term public health policies, and vaccination. High-frequency components, on the other hand, represent rapid changes associated with abrupt weather events, local outbreaks or short-term local interventions. Maximum in PSD spectra and associated frequencies can reveal time scale of main processes affecting influenza spread on annual or daily and decadal scale. The results presented in Section 3.1 clearly demonstrate a relationship between winter duration - determined using the extreme values and inflection points of the seasonality index - and the time scale associated with the maximum in the PSD low-frequency spectra. These findings justify further analysis of the correlation between the influenza infection rate and the seasonality index.

## 2 Material and Methods

2.1. Study area

The global strategy for combining influenza relies on extensive surveillance systems. Under the patronage of the World Health Organization (WHO), member countries collect and analyze data on seasonal influenza trends from hospitals, clinics, and health networks (Hay and McCauley 2018). This international effort forms the backbone of influenza surveillance, enabling the identification of patterns in virus transmission and epidemic peaks. We have collected influenza and climate data from several different European and Asian countries, including Albania, Bangladesh, Estonia, Latvia, Luxembourg, Serbia, Slovenia and Singapore (Fig. 1). Our choice of represented countries was based on several factors: i) duration of influenza data time series - we were aiming to find countries with at least ten years of data; ii) the location of the country – we were seeking data originating from multiple different climate zones; iii) demographic heterogeneity – since influenza data are available on country or regional level and meteorological data on local level, we were seeking for countries, with their capitals, being statistically representative of the entire country. This has given us an excellent overview of the varying dynamics of influenza outbreaks throughout the years over the different regions.

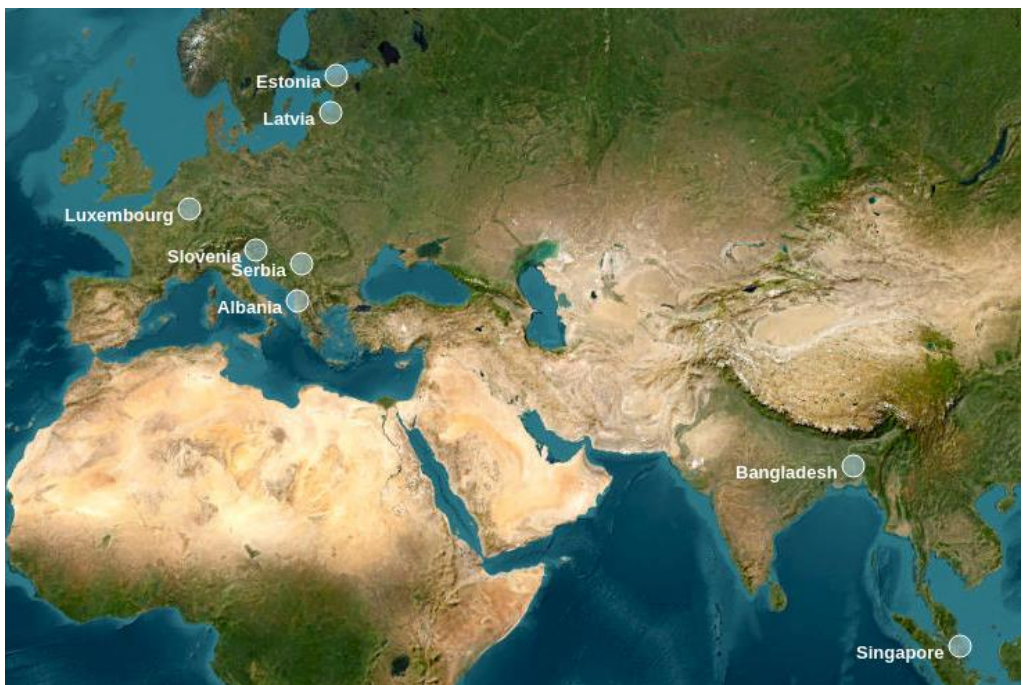

**Figure 1** Geographical distribution of study locations

| Country | Location | Latitude (°), Longitude (°), Altitude (m) | Period | S | NS | ND | Sum |
|---|---|---|---|---|---|---|---|
| Albania | Tirana-la Praka | 41.33; 19.80; 97 | NS: 2015-2016 | 14114 | 9733 | / | 23847 |
| Bangladesh | Dhaka-Tejgaon | 23.77; 90.38; 7 | S: 2019-2020 | 18576 | / | 53535 | 72111 |
| Estonia | Tallinn-Tallinn | 59.41; 24.83; 39 | ND: 2011-2013 | 5212 | 58053 | 6447 | 69712 |
| Latvia | Riga-Riga | 56.96; 24.05; 26 | ND: 2010-2013 | 946 | 49782 | 18032 | 68760 |
| Luxembourg | Luxembourg-Luxembourg city | 49.62; 6.21; 376 | ND: 2010-2013 | 10499 | 2560 | 2372 | 15431 |
| Serbia | Belgrade-Belgrade | 44.80; 20.46; 132 | | 4775 | 5871 | | 10646 |
| Slovenia | Bezigrad-Ljubljana | 46.06; 14.51; 298 | ND: 2010-2013 | 6455 | 196779 | 17154 | 220388 |
| Singapore | Paya Lebar Singapore | 1.36; 103.91; 19 | | / | / | 27517 | 27517 |

**Table 1** Study location and influenza data summary for 2013-2024 period [Legend: S-Sentinel, NS-Non-sentinel, ND-Not defined]

2.2. Influenza data

In this research, we analysed the annual dynamics of the influenza virus spreading using country-based, open-access data governed by the World Health Organisation (WHO). Weekly influenza data included: i) tested individuals, ii) influenza-positive, iii) influenza-negative, and iv) a specific type of influenza-positive. Influenza types recognized by WHO data base include: A(H1), A(H1N1)pdm09, A(H3), A not subtyped, B(Victoria), B(Yamagata), B(linage not determined). The list includes the entire scope of tracked subtypes of influenza viruses, although not every country within this study had a tracking record of all of them. In most cases, A subtypes had a dominant number of infections to B subtypes. Further, data are classified according to surveillance site type as Sentinel, Non-sentinel and Not Defined.

    Non-sentinel surveillance refers to the collection of data originating from a wide array of hospitals and clinics that routinely report findings from influenza testing of outpatients (Murray and Cohen 2016). These tests are often not standardised, meaning they cannot be referenced within the same category as sentinel surveillance. Sentinel surveillance refers to the collection of data obtained

by using standardized testing within predefined institutions, namely laboratories and hospitals that provide a service of testing patients with symptoms potentially connected to influenza infection. These symptoms are often classified as influenza-like-illness (ILI) and severe acute respiratory infections (SARI) (Szecsenyi et al. 1995). It is not uncommon to find significant differences between Sentinel and Non-sentinel influenza time series. The key reason for this division most likely lying in differences in tests performed in each of these sights, as well as the diligence with which the data had been uploaded. Since in non-sentinel influenza environments, this test is just one of many, we cannot be positive that the exact result of the test was accurately published. Other than that, it is also rational to expect a higher number of positive cases in Sentinel testing, exemplifying the essence of these institutions in testing the patients that express ILI or SARI symptoms. However, the potential testing bias is avoided by means of performing standardised tests, regardless of symptoms. Not defined data can be either Sentinel or Non-sentinel. These data often include a lot of noise, inconclusive data entries and gaps in time series, which is why we will omit them in our study.

Naturally, not every country had uniform data within the aforementioned surveillance data types for the exact same period. We identified the seasons from 2013/2014 up to 2023/2024, as a period with best influenza records in all selected countries. In this phase of our research study, we decided to take into account all influence positive cases, but to classify them according to surveillance site type.

2.3. Meteorological data

To integrate meteorological data, SYNOP station data were retrieved using the GSODR library (Sparks 2018; Sparks et al. 2024). The GSODR library in R facilitates access to the Global Surface Summary of the Day (GSOD) dataset, providing easy retrieval and processing of historical daily weather data from NOAA's Integrated Surface Database (ISD) (GSOD.1.0). The dataset includes comprehensive meteorological parameters such as temperature, precipitation, and wind speed, collected from thousands of weather stations worldwide. For selected locations and seasons (Tab. 1) time series of extreme ($T_{max}$, $T_{min}$) and average daily ($T_d$) air temperatures are retrieved and used to calculate seasonality index

$$DTRT = (T_{max} - T_{min})/T_d \tag{1}$$

and its weekly averages. To avoid the seasonality index's high variability, absolute values are used in further analysis.

2.4. Methods

In meteorological practice, particularly when the goal is to identify the relationship between, for example, atmosphere and biosphere phenomena, long-term averages are typically analysed. We also followed this practice in current research, calculating longer-term averages of influenza and meteorological data before further analysis. The main goal is to avoid year-to-year variability of analysed time series, which in the case of influenza data can be a result of data gaps or some other non-influenza related cause (for example, sudden drop in number of positive cases in the middle of influenza season), while in case of meteorological data are typically induced by strong winter and early spring synoptic disturbances. Therefore, we expect to identify the impact of seasonal transition on influenza spread when averages are taken over many years.

Quantitative analysis performed in this study is based on PSD analysis of infection rate time series and correlation between infection rate and seasonality index time series. In order to account for time necessary for infection to spread in the population, lagged and non-lagged correlations are calculated for all locations. While designing seasonal time series for influenza and seasonality index, time as an independent variable is expressed as a pseudo-week aligning the weeks from eight weeks before January 1st to thirty weeks after the start of the year to capture the seasonality.

## 3 Results

To test our hypothesis, we analyzed the characteristics of the infection rate's PSD (Figs. 2–4) and the correlation between infection rate trends and the seasonality index (DTRT) during the influenza season (Figs. 5 - 7).

3.1 Power Spectral Density of Infection Rate Time Series

The PSD analysis was performed using raw infection rate data from various locations and surveillance types (Appendix, Figs. A1–A8). The results show distinct trends across different regions.

*Low-frequency domain*. In European temperate regions (Fig. 2), PSD values gradually decrease keeping low-frequency dominance in time series data, highlighting seasonal or long-term periodicity. Variability in low-frequencies PSD peaks is influenced by climate variability and differences in climate and population behavior among locations. Sentinel data exhibit more consistent and well defined low-frequency peaks as a result of more structured data collection, whereas non-sentinel data shows greater variability.

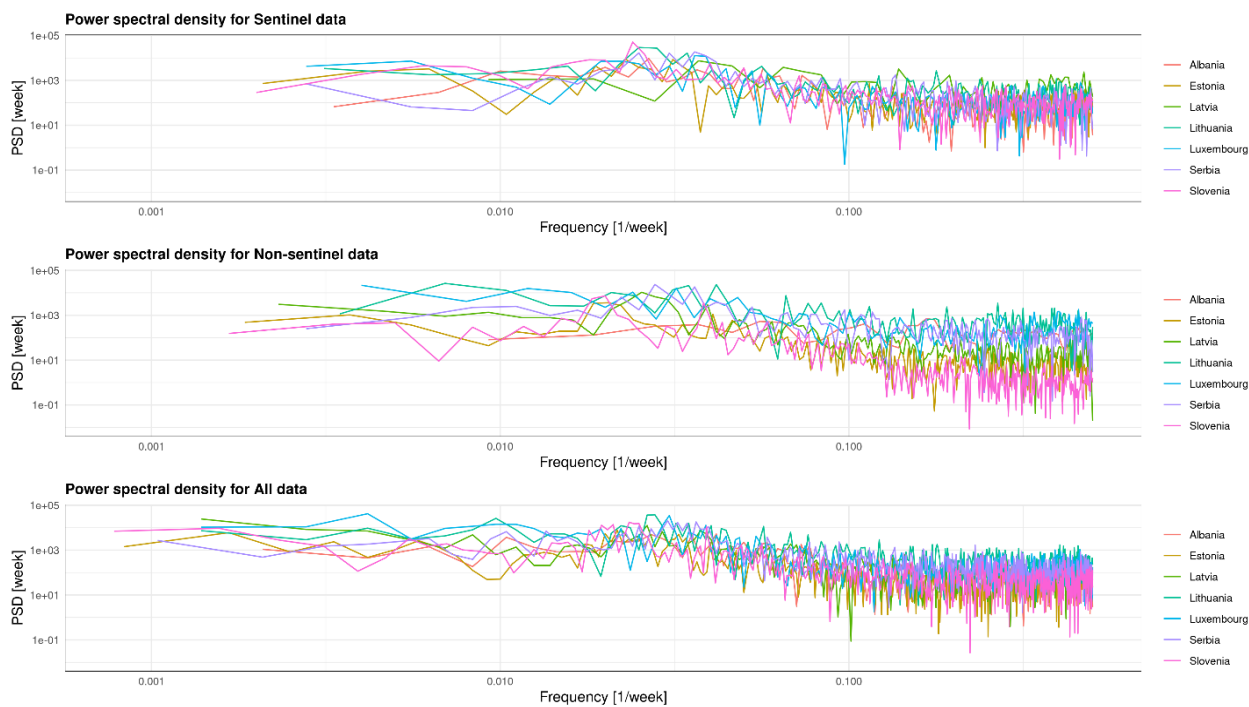

**Figure 2** PSD of infection rate for 2013–2024 influenza seasons for European countries

The gradient between low and high frequencies is less pronounced in Bangladesh's subtropical region and Singapore's tropical region (Fig. 3). PSD peaks are smaller by an order of magnitude, indicating reduced influenza seasonality compared to European temperate regions.

*High-frequency domain.* High-frequency PSD trends reflect short-term events that can affect influenza transmission and could be associated with adverse weather events (snowstorm, excessive rain, for example), local interventions (school closing) or social activities (Christmas-related travelling). These events are cite-specific, leading to variations in high-frequency PSD among European locations. For example, PSD peaks variability is in range from $10^2$ in Sentinel data to $10^4$ in non-sentinel data. In Bangladesh and Singapore PSD variability is up to $10^2$.

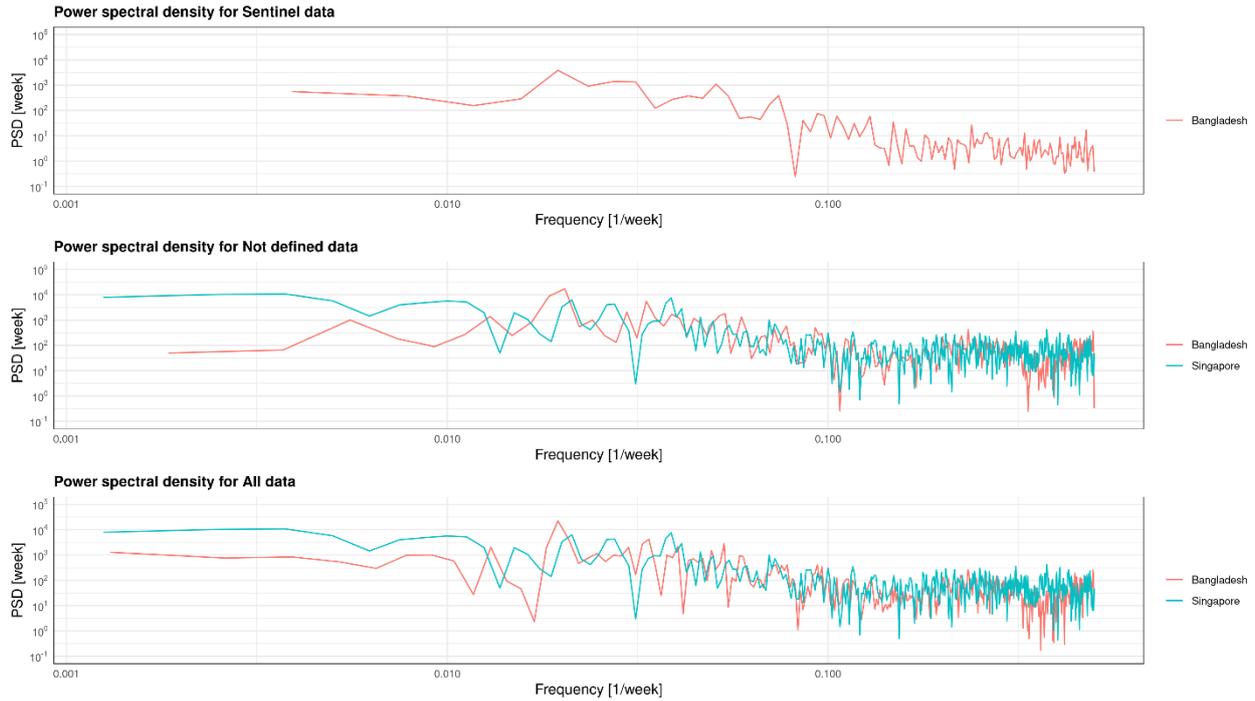

**Figure 3** PSD of infection rate for 2013–2024 influenza seasons in Bangladesh and Singapore

*Correlation with winter duration.* Low-frequency peaks (LFP) and their associated time scales ($\tau$_LFP) differ significantly among locations, likely due to regional climate variations (Tab. 2). In this study, climate characteristics of influenza season are described using seasonality index

| Location | LFP (week$^{-1}$) | HFP (week$^{-1}$) | $\tau$_LFP (week) | $\tau$_HFP (week) | WD (day) |
|---|---|---|---|---|---|
| Albania_S | 0.0267 | 0.1333 | 37.50 | 7.50 | 72 |
| Estonia_NS | 0.0229 | 0.1146 | 43.64 | 8.73 | 151 |
| Latvia_NS | 0.0255 | 0.1296 | 39.27 | 7.71 | 128 |
| Luxembourg_S | 0.0361 | 0.4611 | 27.69 | **2.17** | 55 |
| Serbia_S | 0.0278 | 0.1167 | 36.00 | 8.57 | 50 |
| Serbia_NS | 0.0361 | 0.1361 | 27.69 | 7.35 | 50 |
| Slovenia_S | 0.0200 | 0.1067 | 50.00 | 9.37 | 65 |
| Slovenia_NS | 0.0240 | 0.4160 | 41.67 | **2.40** | 65 |
| Bangladesh_S | 0.0195 | 0.1055 | 51.20 | 9.48 | |
| Singapore_ND | 0.0038 | 0.3750 | - | **2.67** | |

**Table 2** Low- and high-frequency peaks and associated time scales in infection rate PSD for 2013 – 2024 influenza seasons and winter duration (WD).

(DTRT) variation. By comparing winter duration (WD), as defined by Lalic and Firanj Sremac (2025), with $\tau$_LFP we obtained a high correlation (Fig. 4), emphasizing the impact of atmospheric conditions on infection rates. Extremely high values of $\tau$_LFP, particularly for Singapore, indicate lack of seasonal changes. These findings also validate the method for determining winter duration.

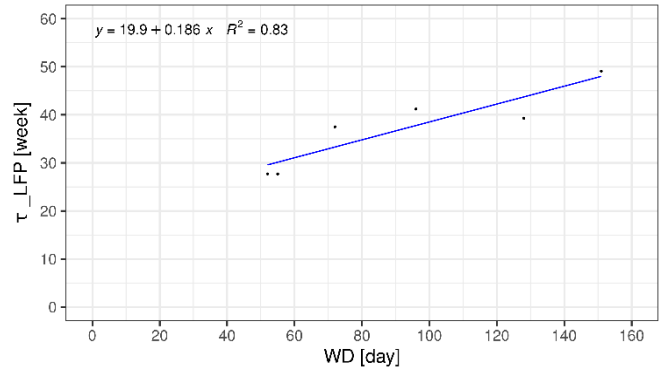

**Figure 4** Duration of winter (WD) compared to LFP time scale ($\tau$_LFP)

*Impact of quality of life*. Notably, the high-frequency time scale ($\tau$_HFP) in Luxembourg, Slovenia and Singapore stands out from all other locations, suggesting a potential link between quality of life and influenza transmission. Namely, according to Numbeo's Quality of Life Index (2024), Luxembourg is ranked as No. 1 and Slovenia as No. 13 in Europe, while Singapore ranks sixth globally. These differences may reflect varying responses to environmental and social factors.

3.2 Correlation between infection rate and seasonality index

To evaluate the relationship between infection rates and the seasonality index, we analyzed the timing of peaks in both time series (Fig. 5; Tab. 3) and their correlations across regions during influenza season (Tab. 4). Since subtropics and tropics are typically considered as regions without exerted influenza seasonality, we calculated correlations over the whole year for Bangladesh and Singapore.

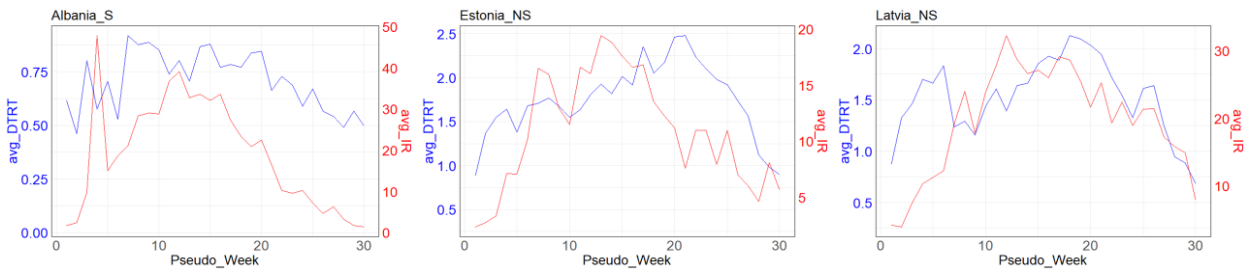

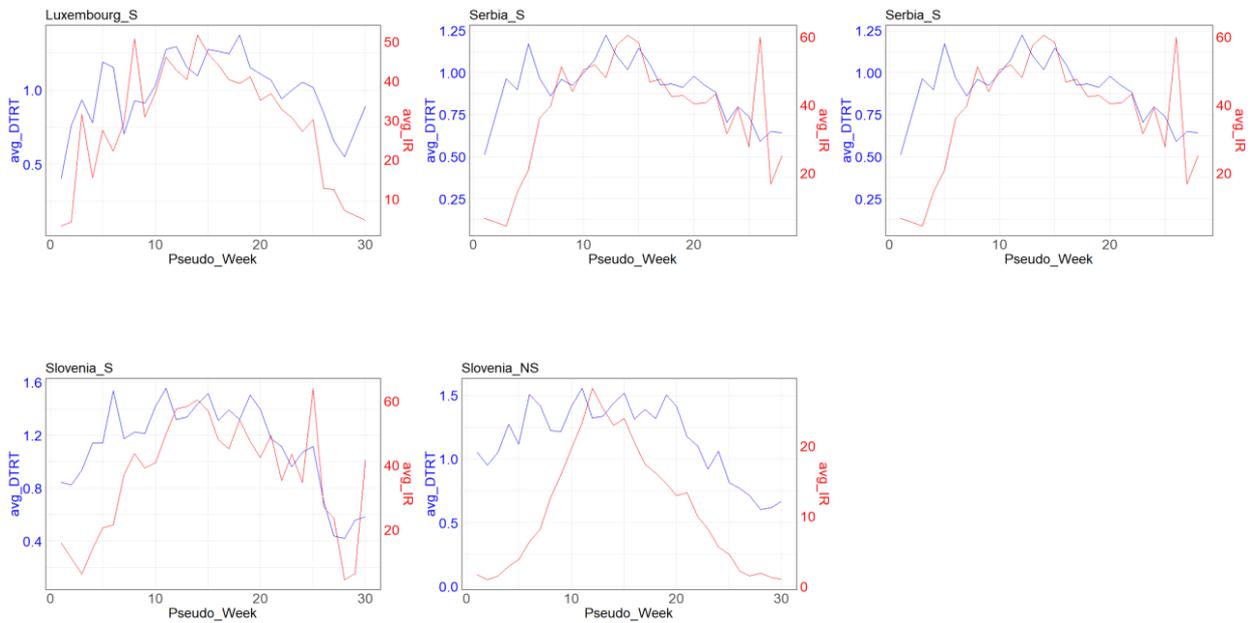

**Figure 5** Variation of infection rate (IR) and seasonality index (DTRT) for 2013 – 2024 influenza seasons in European countries

| Country | Influenza peaks | DTRT peaks | Lag used | Difference |
|---|---|---|---|---|
| Albania_S | [4; 12] | [3; 9] | 1 | [0; 2] |
| Estonia_NS | [7; 13] | [8; 13] | 0 | [-1; 0] |
| Latvia_NS | [8; 12] | [6; 11] | 1 | [1; 0] |
| Luxembourg_S | [3; 5] | [3; 5] | 0 | [0; 0] |
| Serbia_S | [8; 14] | [5; 12] | 2 | [1;0] |
| Serbia_NS | [6; 11] | [5; 11] | 2 | [-1; -2] |
| Slovenia_S | [8; 12] | [6; 11] | 1 | [1;0] |
| Slovenia_NS | [-; 12] | [6; 11] | 1 | [-; 0] |

**Table 3** Timing of first and secondary peaks in time series and difference among peaks after lagging. [Difference is calculated as (influenza peak – lagged DTRT peak)]

*Timing of peaks.* In temperate regions, infection rate peaks typically occur between pseudo-weeks 8 and 14, while seasonality index peaks occur earlier, generally between pseudo-weeks 3 and 12. Influenza infection rate peaks generally lag behind seasonality index peaks by 0–2 weeks, implying that in some regions, it is important to account for a potential time lag between environmental conditions and the first symptoms. Therefore, correlations between time series are calculated for both lagged and non-lagged correlations, using p-value as a measure of correlations (Tab. 4; Fig. 6).

| Country | r_non-lagged | p-value | r_lagged | p-value |
|---|---|---|---|---|
| Albania_S | 0.595 | $5 \cdot 10^{-4}$ | **0.721** | $9 \cdot 10^{-6}$ |
| Estonia_NS | **0.511** | $3 \cdot 10^{-3}$ | - | - |
| Latvia_NS | **0.542** | $2 \cdot 10^{-3}$ | 0.51 | $5 \cdot 10^{-3}$ |
| Luxembourg_S | **0.763** | $1 \cdot 10^{-6}$ | - | - |
| Serbia_S | 0.425 | $3 \cdot 10^{-2}$ | **0.674** | $2 \cdot 10^{-4}$ |
| Serbia_NS | 0.566 | $2 \cdot 10^{-3}$ | **0.670** | $2 \cdot 10^{-4}$ |
| Slovenia_S | 0.636 | $1 \cdot 10^{-4}$ | **0.721** | $1 \cdot 10^{-5}$ |
| Slovenia_NS | 0.736 | 3·10⁻⁶ | **0.774** □ | 8·10⁻⁷ |

**Table 4** The overview of the correlation between the infection rate (IR) and seasonality index (DTRT) for 2013 – 2024 influenza seasons in European countries [Crosses in superscript indicate a number of lagged weeks.].

*Correlation analysis*. In Latvia, introducing a one-week lag based on peak shifts failed to capture seasonal trends, resulting in decreased correlation. In Serbia_NS, on the other hand, lags that correlated well with seasonal trends missed infection rate peaks. This raises the dilemma of whether it is more important to predict influenza peaks or overall seasonal trends. In contrast, lags based on the first two peaks brought high correlations and good alignment with at least one peak in Luxembourg_S, Serbia_S, Slovenia (S and NS) and Albania_S. In Serbia's sentinel data, a two-week lag significantly improved the correlation from 0.425 to 0.674 and increased statistical significance (p-value from 3·10⁻² to 2·10⁻⁴). Similar correlation improvements are obtained with one-week lags in Albania and Slovenia, where Slovenia's sentinel data exhibited particularly notable changes.

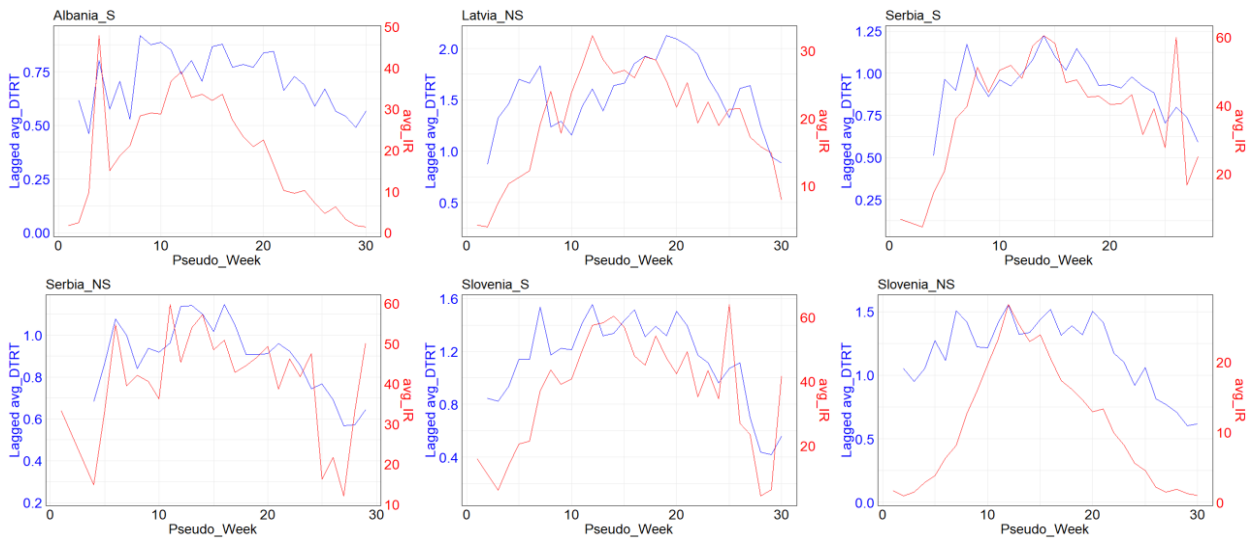

**Figure 6** Variation of infection rate (IR) and lagged seasonality index (DTRT) for 2013 – 2024 influenza seasons in European countries.

The seasonal strength of the DTRT index on Bangladesh and Singapore locations is in accordance with their respective climate characteristics - high (0.82) and medium (0.43). Extremely low variability of DTRT in Singapore witnesses about its Equatorial climate with low annual variation of seasonal drivers, while more exerted one in Bangladesh is a result of monsoon driven subtropical climate.

In Bangladesh, the seasonal strength of influenza infection rate time series is medium (0.36) with one maximum during the year. A highly negative correlation between IR and DTR over the year (r = -0.685; p = 2·10⁻⁸) is kept in one-week lagged correlations (r = -0.646, p = 3·10⁻⁶). Even though the IR seasonal strength index is very low for Singapore (0.076), two distinct peaks could be identified. The first, at the beginning of the year (first two weeks), is most probably related to travelling from Christmas through New Year. The second peak, from May to early August, could be associated with a dengue outbreak, causing intensive testing and an increase in the number of influenza-positive cases. Namely, although DTRT varies by a narrow margin, Fig. 7 shows the correlation between IR second peak and summer-specific annual minimum of DTRT. Moreover, the 5-week lagged correlation between IR and DTRT time series is **0.985** with a p-value of **7·10⁻⁴**. A possible scenario is that weather conditions associated with DTRT drop are favorable for mosquito population development, which will, in several weeks (possibly 5 weeks in our case), cause an increase in dengue infection cases.

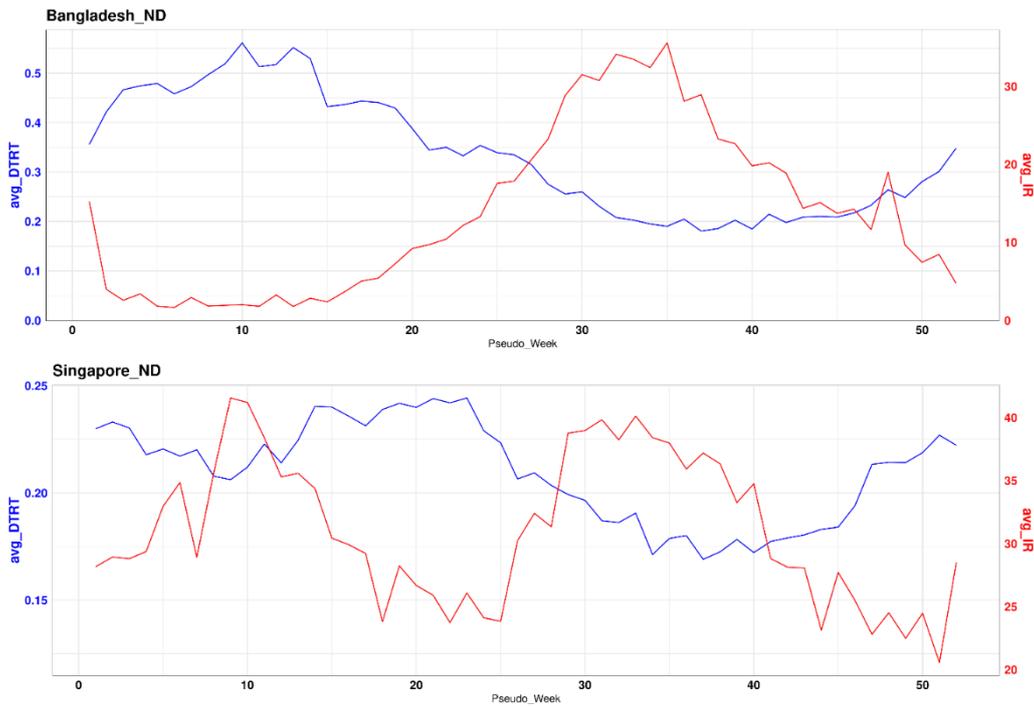

**Figure 7** Variation of infection rate (IR) and seasonality index (DTRT) for 2013 – 2024 influenza seasons in Bangladesh and Singapore

## 4 Discussions

This study aims to enhance our understanding of influenza seasonality in temperate regions by analyzing the relationship between infection rates and seasonal transition. Using novel seasonality index (DTRT) and PSD analysis, we explored how their use could enhance predictions of influenza outbreaks and inform public health interventions.

A key finding of this study is the high correlation between infection rates and the seasonality index. The DTRT effectively captured the seasonal dynamics influencing influenza transmission. The improved correlations obtained when using lagged time series underscore the time-dependent relationship between atmospheric seasonality and infection dynamics (Shaman et al. 2010; Tamerius et al. 2011). The high (negative) correlations obtained for locations in Bangladesh and Singapore indicate a new avenue for influenza prediction in subtropical and tropical regions.

Regional differences in the shifts between DTRT and influenza rate peaks highlight the influence of social and behavioral factors. One of the most notable mechanisms of influenza spreading within a population is through age-specific transmission, leading to potentially more than one peak. This trend was observed in Serbia, where there are usually two peaks of influenza infections, one associated with New Year's Eve and the other after Orthodox Christmas. The second peak, occurring in January and February, is most probably related to children returning to school, which leads to another wave of influenza transmission before the end of the cold period in these temperate regions.

Social and behavioral factors introduce a new layer of uncertainty in our conclusions related to data collection and reporting. Namely, the time period that separates a new data point in the database, and the infection itself is divided by a period of incubation of the pathogen within the body, before expressing recognizable symptoms and behavioral framework, i.e. the process of decision-making to get tested, the lack of diligence in uploading the test results in the testing facility and so on. Studies addressing behavioral aspect of the influenza pandemics are scarce however, as shown by Peppa M et al. (2017), it seems as though the severity of ILI-like symptoms directly correlate to decision-making aspect of getting tested, or in this case, reporting the experienced symptoms. Subsequently, it also impacts surveillance reports and classification methods, which may lead to differently structured datasets than expected.

Another significant finding is the high correlation between PSD low-frequency peaks and the duration of winter, as determined by the extreme values and inflection points of DTRT time series. This confirms that long-term atmospheric patterns, such as winter duration, significantly influence influenza transmission. PSD analysis shows low-frequency dominance in temperate regions, emphasizing the role of accurate determination of season transitions in reliable assessment of infection trends, which consequently imply public health preparedness. The ability to predict influenza peaks using seasonality index forecast, based on numerical weather prediction, could flatten the curve of influenza through early warning system and prevention measures. This could help to reduce mortality rates and healthcare pressures during epidemics (Ryu and Cowling, 2020).

Key study limitations are related to variability in data collection methods, particularly between sentinel and non-sentinel data, introducing potential biases. Additionally, behavioural studies indicate that symptoms severity directly impacts decisions to seek testing, influencing the structure and reliability of surveillance data (Kim et al., 2017).

Our future research will take several avenues.

i. Testing the efficacy of extended (10 days), monthly, and seasonal weather forecasts in forecasting influenza transition dynamics; assessing the potential impact of climate change on regional diversity.
ii. Improvement of SIR model parameterisations using seasonality index.
iii. Incorporation of social and behavioural factors in SIR model using Physics Informed Neural Networks (PINN) following methodology described by Cuong et al. (2024).
iv. Hybridisation of SIR model in order to incorporate social, behavioural, and epidemiological factors, meaning transmissibility and survivability of the virus, resulting in a higher infection rate, while keeping explainability of physics-based (SIR) model.

# 5 Conclusions

Influenza spread and seasonal changes are complex phenomena, challenged further by climate change unfolding at an unprecedented rate. Atmospheric conditions affect influenza transmission directly and indirectly on different time scales. This study presents the methodology based on seasonality index dynamics and PSD analysis of influence time series to predict influenza peaks and overall dynamics to improve public health preparedness. By bridging the gap between climate science and epidemiology, we provide robust framework for improved understanding of influenza seasonality. As climate change globally reshapes atmospheric, and behavioral patterns as well as seasons duration, such transdisciplinary approach is essential for better understanding of influenza transmission as well as the spread and mitigation of its global health impact.

**Acknowledgement**

This research is supported by the European Union (Grant Agreement No. 101136578). The views and opinions expressed in this publication are solely those of the author(s) and do not necessarily represent the official position of the European Union or the European Commission. Neither the European Union nor the granting authority is responsible for any use that may be made of the information contained herein. Additional support was provided by the Ministry of Science, Technological Development, and Innovation of the Republic of Serbia through two Grant Agreements with the University of Novi Sad, Faculty of Agriculture (No. 451-03-66/2024- 03/200117, dated February 5, 2024).

**Appendix**

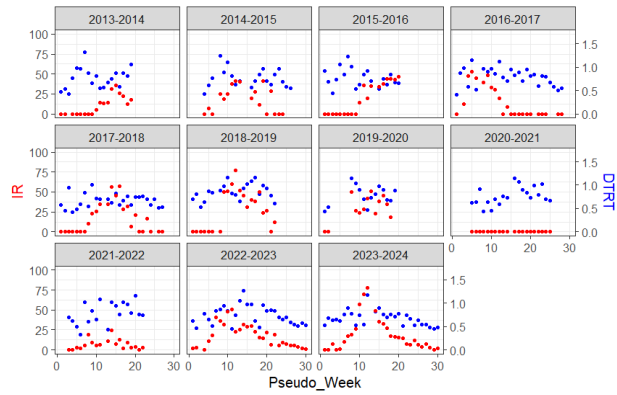

**Figure A1** Annual variation of influenza infection rate in Albania for sentinel data

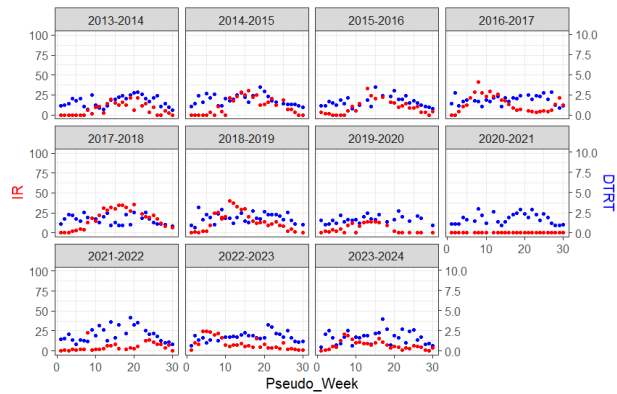

**Figure A2** Annual variation of influenza infection rate in Estonia for non-sentinel data

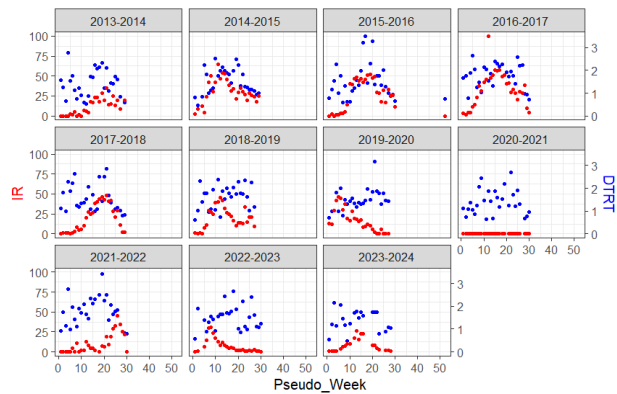

**Figure A3** Annual variation of influenza infection rate in Latvia for non-sentinel data

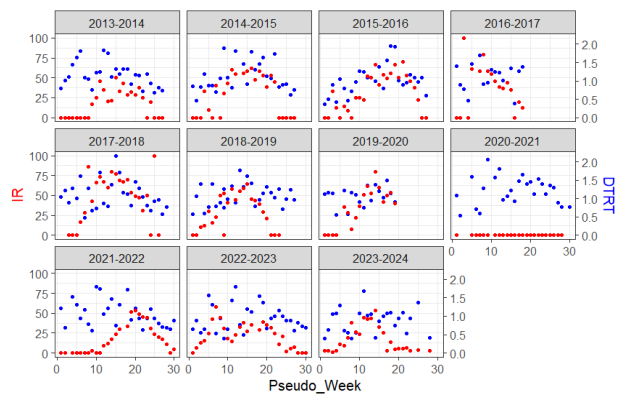

**Figure A4** Annual variation of influenza infection rate in Luxembourg for sentinel data

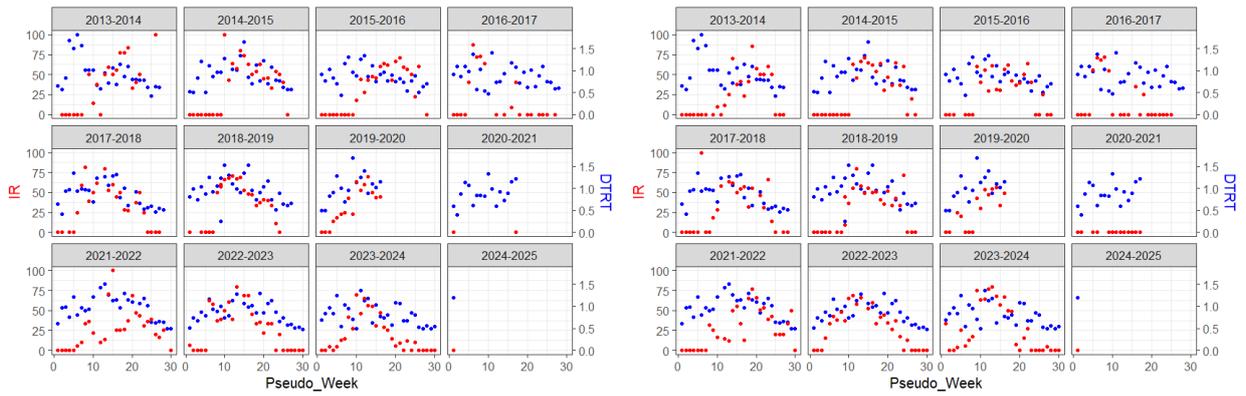

**Figure A5** Annual variation of influenza infection rate in Serbia for sentinel data (left) and non-sentinel data (right)

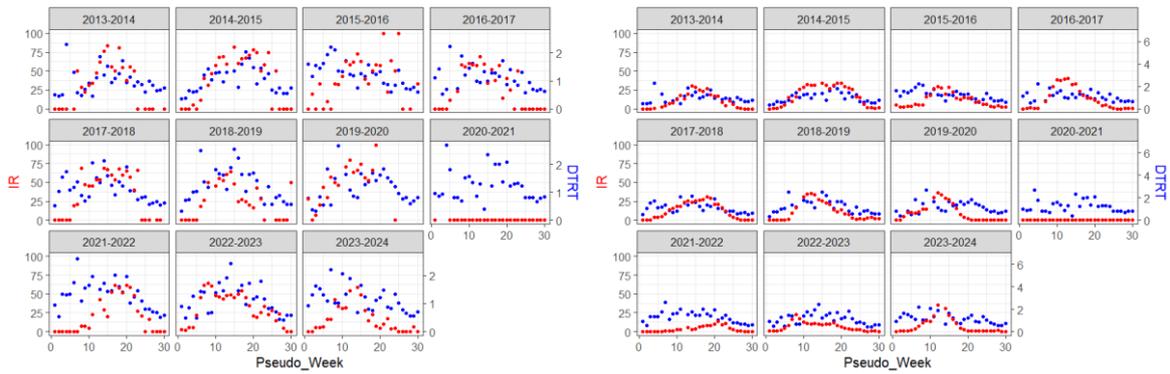

**Figure A6** Annual variation of influenza infection rate in Slovenia for sentinel data (left) and non-sentinel data (right)

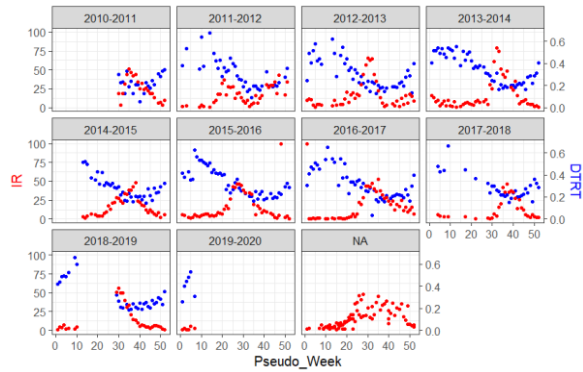

**Figure A7** Annual variation of influenza infection rate in Bangladesh for non-defined data

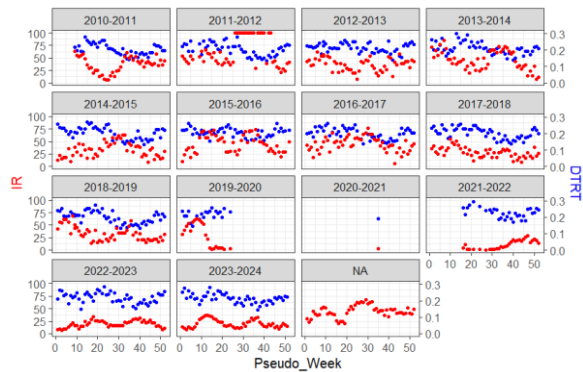

**Figure A8** Annual variation of influenza infection rate in Singapor for non-defined data